# Thermal response of the iron-based Ba122 superconductor to *in situ* and *ex situ* processes

Shinnosuke Tokuta[1] and Akiyasu Yamamoto[1]

[1]Department of Applied Physics, Tokyo University of Agriculture and Technology, Koganei, Tokyo 184-8588, Japan

E-mail: s195941r@st.go.tuat.ac.jp

**Abstract**

The thermal properties are one of the key parameters to control phase purity and microstructure of polycrystalline materials. The melting point of the iron-based $BaFe_2As_2$ superconductor (Ba122), which foresees high-field applications, remains controversial. In this work, thermogravimetry-differential scanning calorimetry measurements (TG-DSC) of undoped and Co-doped Ba122 were carried out. Mixtures of elemental metals and pre-reacted Ba122 powders were prepared to investigate the thermal responses during *in situ* and *ex situ* synthesis routes, respectively. In addition, the phases and microstructures of the quenched samples were evaluated to elucidate the observed exothermic/endothermic peaks. Our results suggest that the melting point of Ba122 is ~1300°C.

# 1. Introduction

BaFe$_2$As$_2$ (Ba122) is a promising material for high magnetic field applications among the iron-based superconductors [1, 2] owing to its high transition temperature ($T_c$), high upper critical field with low electromagnetic anisotropy [3-8], and large critical grain boundary angle [9, 10]. In the last decade, efforts have been made to develop superior critical current performance in Ba122 wires [11, 12], bulks [13, 14] and thin films [15]. There are several factors that should be overcome to improve critical current performance of iron-based superconductors, such as intrinsic weak-link [9, 10, 16], formation of impurities due to relatively reactive elements in its constituent elements, and amorphous/impurity phases and compositional fluctuations at grain boundaries [17-21]. The heat treatment conditions are one of the parameters that affect these factors. For synthesis of Ba122 polycrystalline materials, *in situ* methods using elemental metals and/or precursor alloys as starting materials and *ex situ* methods using pre-reacted Ba122 as starting material are frequently used, and heat treatment temperatures are typically 600–1100°C [11-14, 22-25]. It is important to clarify the thermal properties of Ba122, such as the melting point, for control of phase formation and microstructure.

Since the melting point is important for determining the growth conditions of single crystal, it has been reported mainly in the process of crystal growth research. A flux method is often used to grow Ba122 single-crystals, and heat treatment temperatures are 980–1000°C with the Sn-flux [26, 27] and 1000–1200°C with the self-flux [5, 26, 28-31]. Although there are several reports measuring thermal responses of Ba122, the data interpretation varies between groups and the melting point remains controversial. Morinaga *et al.* established a pseudo-binary phase diagram for Ba122 and FeAs based on the heat treatments and elemental analyses of Ba122 or mixtures of Ba122 and FeAs with different molar ratios and reported that Ba122 melts congruently above 1170°C [32]. Peng *et al.* measured the thermal response of Ba122 powder; K28%-doped Ba122 powder; and Ba, Fe, and As mixtures in a Sn flux up to ~1000°C using differential thermal analysis (DTA) and reported that Ba122 and K28%-doped Ba122 melt incongruently at 880°C and 832°C, respectively, and Ba122 forms at 591–647°C and crystallizes at 825–913°C [33]. Sun *et al.* investigated Ba122 and K28%-doped Ba122 single-crystal prepared by the Sn flux method using TG-DTA and reported that they melt incongruently at 1030°C and 917°C, respectively [26].

In this study, the thermal responses of Ba122 and Co-doped Ba122 up to 1300°C were measured using TG-DSC for mixtures of elemental metals and pre-reacted Ba122 powders corresponding to *in situ* reaction and *ex situ* synthesis, respectively. In addition, the phase compositions and microstructures of the samples quenched at selected temperatures were evaluated using X-ray diffraction (XRD), scanning electron microscope (SEM), and energy-dispersive X-ray spectroscopy (EDX) in order to elucidate the formation temperatures and melting points of Ba122 and impurities through comparison with TG-DSC results.

## 2. Experimental details

The TG-DSC measurements were performed on four types of powders, which were prepared according to the following procedures. In an Ar glove box, two mixtures of Ba (chunk, 3N), Fe (150 μm, 3N), Co (1–3 μm, 3N), and As (0.5–1 mm, 6N) were prepared, with molar ratios of Ba:Fe:As = 1:2:2 and Ba:Fe:Co:As = 1:1.84:0.16:2, respectively. The mixtures were reasonably ground using a planetary ball-mill apparatus (Fritsch, Premium line P-7, zirconia jar and balls) at 200RPM for 40 h to prepare undoped metal mixture (U-MM) and Co-doped metal mixture (Co-MM) powders. Figure 1 shows the powder XRD patterns of the TG-DSC samples. The XRD patterns of U-MM and Co-MM powders showed Fe and background (corundum, holder) peaks and no Ba122 peaks. Therefore, these powders are suitable for *in situ* reaction to form Ba122. Mechanically alloyed undoped and Co-doped Ba122 powders were prepared by ball-milling of elemental metals with the same molar ratios at 800–1000RPM for ~1 h [25, 34]. The powders were pressed into pellets with diameters of 7 mm, vacuum sealed in quartz tubes, and heat-treated at 600°C for 48 h to obtain well crystallized Ba122 phase. The heated pellets were crushed to prepare undoped Ba122 (U-Ba122) and Co-doped Ba122 (Co-Ba122) powders. The XRD patterns of these powders showed nearly single-phase Ba122 peaks owing to mechanical alloying [25, 34]. Therefore, these powders are suitable for *ex situ* synthesis route. Transport and magnetic measurements confirmed that $T_c$ of the Co-Ba122 sample is 26.3 K [25, 34].

The TG and DSC were simultaneously carried out using the STA 449 F3 Jupiter (NETZSCH). The mass change of a sample associated with vaporization, oxidation, reduction, and so on was measured by TG. The heat generation and absorption of the sample associated with the state change were detected by DSC. Approximately 8 mg of the sample powder was placed in a PtRh container and TG-DSC was carried out in an Ar atmosphere (200 mL/min). The temperature was held at 40°C for 10 min and then increased to ~1300°C at a rate of +20°C/min. To identify the reactions occurring at DSC peak temperatures, Co-MM pellets were vacuum sealed in quartz tubes, heated to 280, 420, 550, 700, and 900°C at a rate of +20°C/min, and quenched. Based on the same procedure, a sample was prepared using U-MM (quenched at 1000°C). The phase composition was determined using powder XRD (Bruker, D2 PHASER, Cu-$K_\alpha$ with $\lambda$ of 1.5418Å) and microstructural observations and elemental analyses were carried out using SEM (Hitachi High-Tech, S-3400N) and EDX (Bruker, QUANTAX, XFlash), respectively.

## 3. Results

The TG curves are shown in figure 2 (a). Little change was observed up to 1300°C in all samples. This indicates that the vaporization, oxidation, or reduction are insignificant during the TG-DSC measurements. The DSC curves are shown in figure 2 (b). The DSC curves of the undoped or Co-doped samples were insignificantly different, while the curves obtained for the *in situ* and *ex situ* samples were significantly different. The exothermic and endothermic reaction were detected as the positive and negative peaks in figure 2 (b), respectively. The DSC curves of the U-Ba122 and Co-Ba122 *ex situ* samples showed no exothermic peaks. On the other hand, the DSC curves of the *in situ* samples showed exothermic peaks at 350, 380, 640, and 720°C for U-MM and 350 and 630°C for Co-MM. These peaks are considered to be due to the formation of Ba122 and impurities under the conditions of this study. Regarding the endothermic reaction, peaks can be observed at ~930°C (except for U-Ba122) and ~1010 and ~1300°C in all samples. In addition, endothermic peaks can be observed at 850 and 950°C in U-MM and 850 and 1100°C in Co-MM.

To identify the reactions that occurred at the DSC peaks, quenched samples were prepared. Figure 3 shows the powder XRD patterns of the samples quenched from selected temperatures. The FeAs, broad Ba122, and sharp Ba122 peaks started to appear in the samples quenched at 280°C, 420°C, and 700°C, respectively. Figure 4 shows the backscattered electron images of the polished surfaces of the samples quenched at (a) 280, (b) 420, (c) 700, and (d) 1000°C. The analysis of the chemical composition using EDX revealed that the dark gray contrast in figure 4 (a) corresponded to Fe and the gray contrast corresponded to the Ba–(Fe,Co)–As phase with a molar ratio of Ba:(Fe,Co):As

= 12:15:30. A few micrometers of Fe particles was dispersed in a matrix that contains less Fe than Ba122, suggesting that this was the microstructure before the chemical reaction occurred. The black, gray, and white contrasts in figure 4 (b) correspond to Fe, FeAs, and the ternary Ba–(Fe,Co)–As phase (molar ratio of 12:18:33), respectively. Although Ba122 did not form, FeAs formed around the Fe particles. The gray and white contrasts in figure 4 (c) correspond to Ba122 and oxidized impurities, respectively, and Ba122 was observed as the main phase. The dark gray and gray contrasts in figure 4 (d) correspond to $Fe_2As$ and Ba122, respectively. A wetting $Fe_2As$ phase can be observed in the gaps between the rectangular-shaped Ba122 grains. The Ba122 grains maintained their shapes, indicating that Ba122 did not melt below 1000°C. However, it is possible that Ba122 grains melted incongruently and $Fe_2As$ phase were formed at 700-1000°C, since there were no large amounts of $Fe_2As$ in the samples quenched at 700°C or below.

## 4. Discussion

Based on the above-mentioned results, the reactions corresponding to the DSC peaks in figure 2 (b) were identified. Table 1 summarizes the DSC peak temperatures for each sample and the corresponding reactions. The exothermic peak at 350°C during *in situ* heating corresponds to the formation of FeAs based on the comparison between the XRD patterns (figure 3) and SEM images (figure 4 (a) and (b)). Similarly, the exothermic peaks at 630 and 640°C correspond to the formation of Ba122 under the conditions of this study. The exothermic peaks at 380 and 720°C observed in U-MM have not been identified. The formation temperature of Ba122 is roughly in agreement with the value reported by Peng *et al.*, that is, 590–650°C in a Sn flux [33]. In figure 3, the broad Ba122 peaks were observed even in the sample quenched at 420°C, which is presumably because the slight mechanical alloying occurred during the preparation of Co-MM. The formation peak of Ba122 in U-MM was smaller than that of Co-MM, which is likely due to the higher amount of mechanically alloyed Ba122 in U-MM compared with Co-MM. With respect to the endothermic reactions, the melting points of $FeAs_2$, FeAs, and $Fe_2As$ are 1020, 1030–1040, and 930°C, respectively, and 840–850°C is the eutectic point of Fe and $Fe_2As$ based on the As–Fe binary phase diagram [35, 36]. These temperatures are roughly consistent with the endothermic peak temperatures of 850, 930, and 1010°C in figure 2 (b). The relatively strong endothermic peak at 1300°C is considered to correspond with the melting of Ba122 because this is the only endothermic peak clearly observable in the *ex situ* samples; the As–Fe and As–Co binary phase diagrams do not contain a corresponding melting point [35-37]. This result is consistent with that of Morinaga *et* *al.* who reported that the melting point of undoped Ba122 is >1170°C [32]. However, our value is considerably higher than the melting points reported by Peng *et al.* and Sun *et al.*, that is, 880°C [33] and 1030°C [26], respectively. The peak at 950°C observed in U-MM can be also observed in other Co-doped powders. The 950 and 1100°C (Co-MM) peaks remain unidentified.

Here, we briefly discuss the formation mechanism of Ba122 under the conditions of this study. Generally, the formation temperature varies depending on the reaction route and the condition (reactivity, size, spatial distribution, and so on) of reactants. The sudden increase of DSC curves observed at ~630°C is likely due to that fine and uniformly dispersed, partially mechanically alloyed reactants simultaneously turned into Ba122. Another possibility is that liquid phase existed and assisted the rapid formation of Ba122, since solid state diffusion coefficients are not so large. In this scenario, the difference in the intensity of peaks at ~630°C between U-MM and Co-MM is considered to be due to the change in the formation of the liquid phase and diffusion coefficient depending on the presence of Co. Though residues of the liquid phase was not observed by SEM and endothermic peak due to melting was not detected by DSC below 630°C, it is interesting to clarify whether such liquid phase exists and what its composition is, for example by quenching experiments at ~630°C.

## 5. Conclusion

In this study, TG-DSC measurements up to high temperature of 1300°C were carried out on *in situ* and *ex situ* undoped Ba122 and Co8%-doped Ba122 powders. The TG curves showed that no significant vaporization, oxidation, and reduction occurred during the measurements. In the DSC curves, endothermic peaks were found in all samples at similar temperatures, and exothermic peaks were found only in the metal mixtures, *i.e. in situ* reactions. The comparison with the data obtained from XRD, SEM, and EDX revealed that FeAs and Ba122 formed at ~350 and 630–640°C, respectively, under the condition of this study. Based on the experiments, the melting point of Ba122 is considered to be ~1300°C.

**Acknowledgments**
The authors would like to thank Mr. Takehiro Kajiwara (NETZSCH Japan K.K.) for their help with the TG-DSC measurements. This work was supported by JST CREST (JPMJCR18J4), JSPS KAKENHI (JP18H01699) and by MEXT Elements Strategy Initiative to Form Core Research Center (JPMXP0112101001).6

**Figure Captions**

Figure 1.

Powder XRD patterns of the samples. U-MM: undoped metal mixture, Co-MM: Co8%-doped metal mixture, U-Ba122: undoped Ba122 powder, and Co-Ba122: Co8%-doped Ba122 powder.

Figure 2.

(a) Mass change of the sample with increasing temperature (thermogravimetry, TG) and (b) thermal reaction of the sample with increasing temperature (differential scanning calorimetry, DSC) with corrected offset.

Figure 3.

Powder XRD patterns of quenched samples prepared using the Co8%-doped metal mixture (280–900°C) and undoped metal mixture (1000°C).

Figure 4.

Backscattered electron images of the polished surfaces of quenched samples prepared using the Co8%-doped metal mixture [(a) 280, (b) 420, (c) 700°C] and undoped metal mixture [(d) 1000°C].

Table 1.

Temperatures of the DSC peaks observed in each sample and corresponding reactions under the conditions of this study.

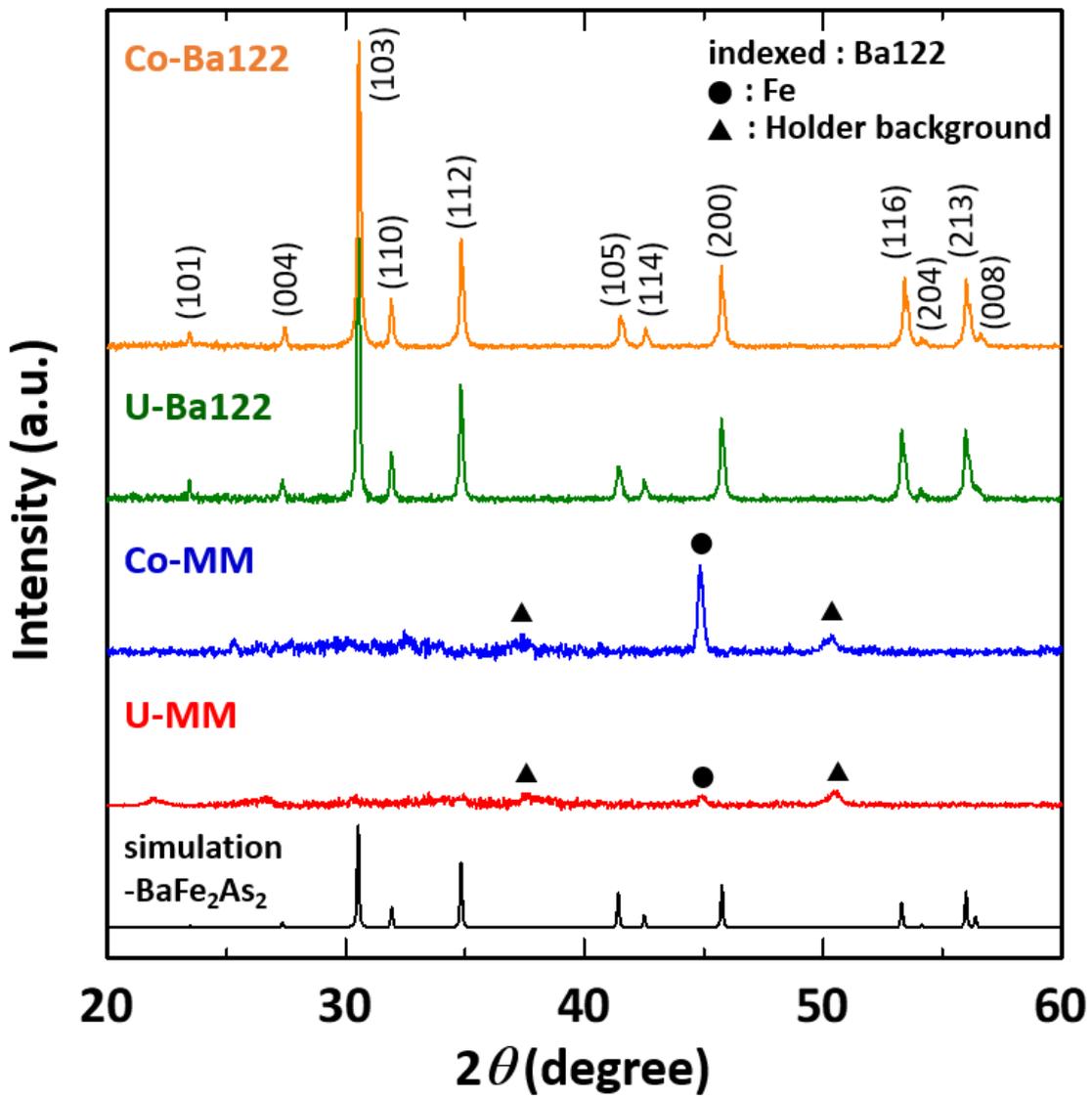

Figure 1.

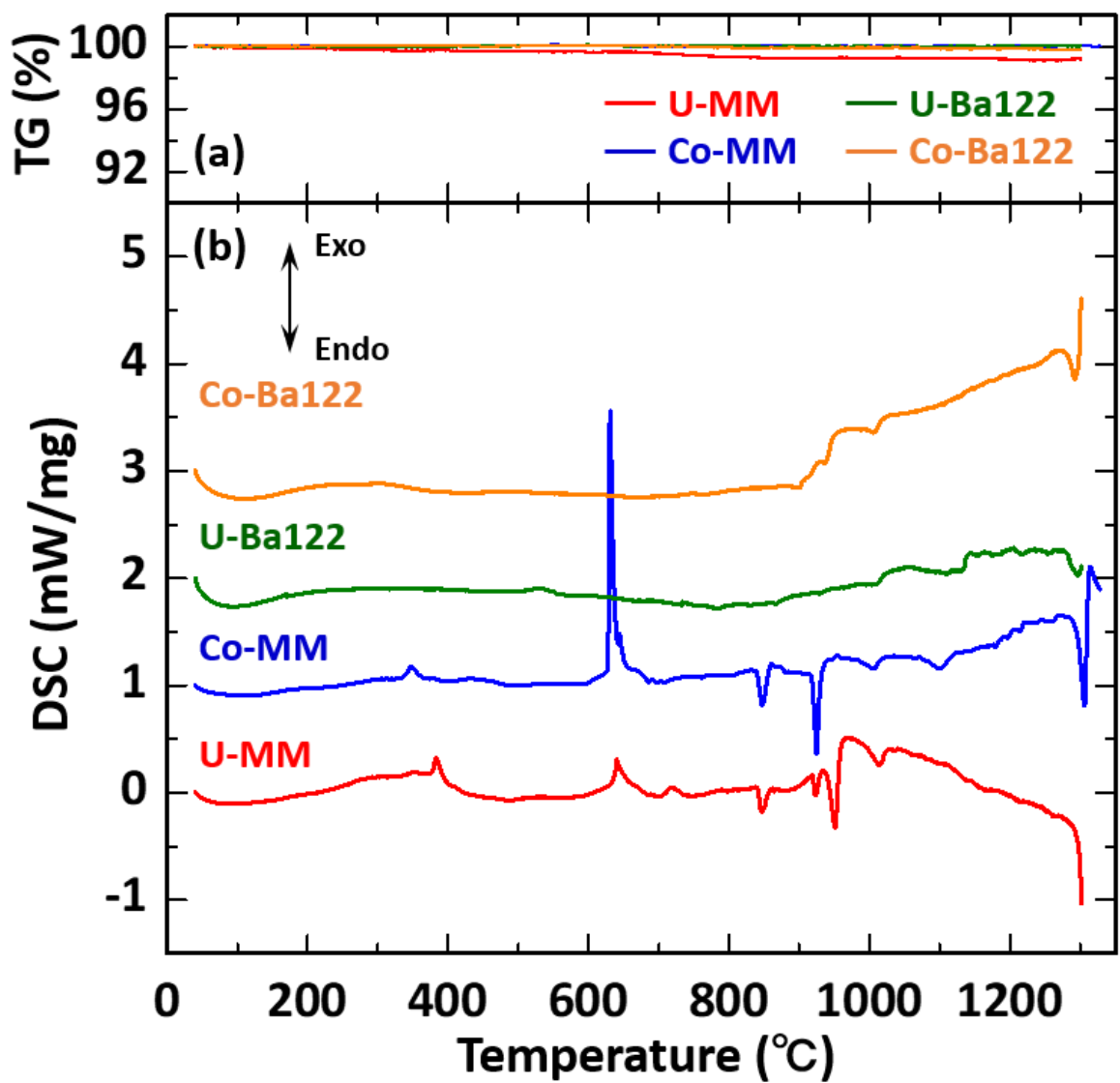

Figure 2.

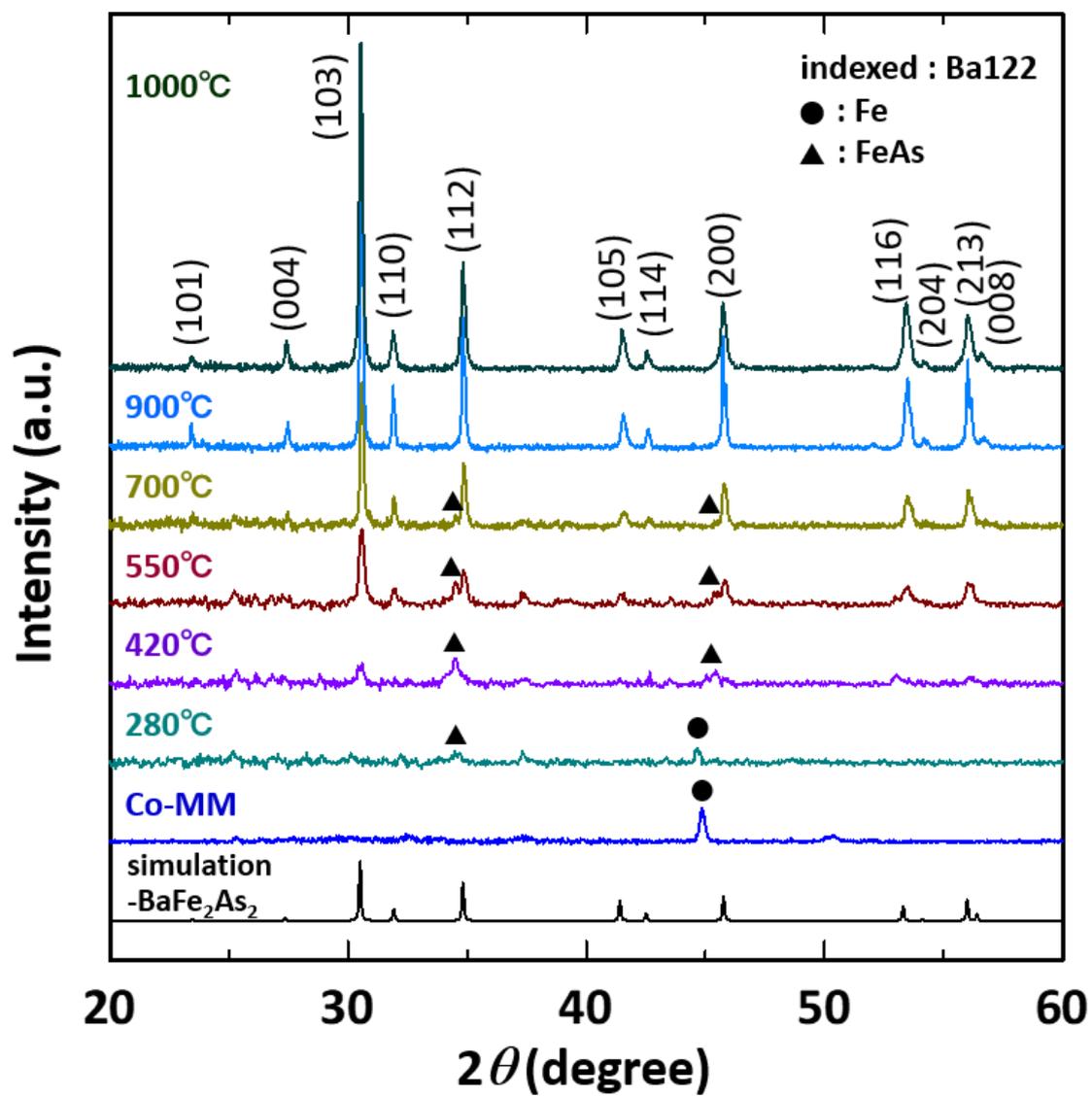

Figure 3.

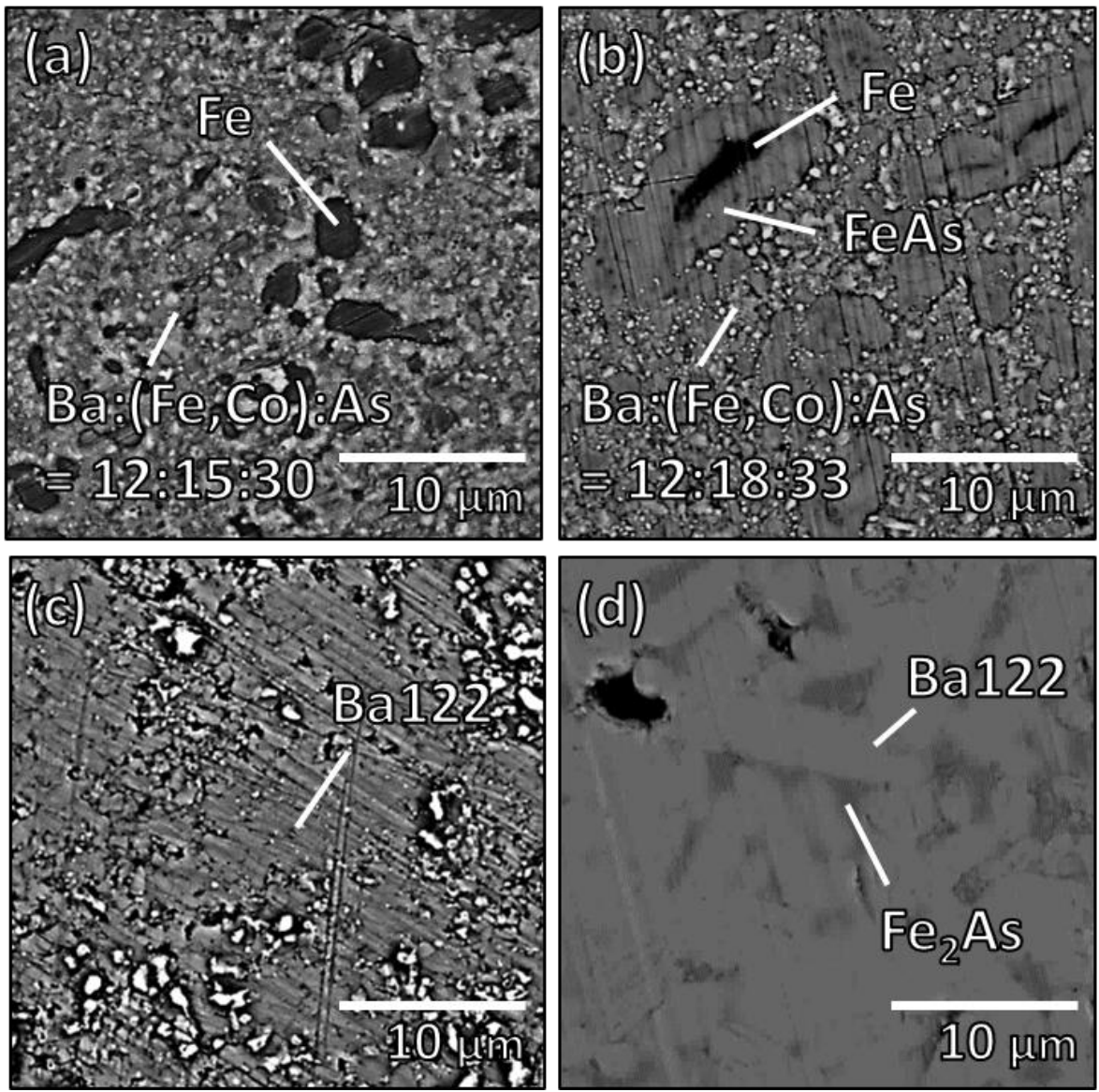

Figure 4.

Table 1.

|  | U-MM | Co-MM | U-Ba122 | Co-Ba122 | Reaction |
|---|---|---|---|---|---|
| Exothermic peaks | 350°C | 348°C |  |  | Formation FeAs |
|  | 383°C |  |  |  |  |
|  | 640°C | 630°C |  |  | Formation Ba122 |
|  | 719°C |  |  |  |  |
| Endothermic peaks | 847°C | 847°C |  |  | Melting Fe and $Fe_2As$ |
|  | 922°C | 924°C |  | 936°C | Melting $Fe_2As$ |
|  | 951°C |  |  |  |  |
|  | 1014°C | 1007°C | 1007°C | 1007°C | Melting FeAs or $FeAs_2$ |
|  |  | 1099°C |  |  |  |
|  | ~1300°C | 1306°C | 1298°C | 1293°C | Melting Ba122 |